\documentstyle[epsfig,11pt]{article}
\def\be{\begin{equation}}
\def\ee{\end{equation}}
\def\al{\alpha}
\def\bt{\beta}
\def\ld{\lambda}
\def\sg{\sigma}

\def\bk{\!\!\!/}

\begin{document}
\thispagestyle{empty}
\setcounter{page}0

~\vfill
\begin{center}
{\Large \bf A new mass relation among the hadron vector resonances.}\\

\vspace{.5cm}
{\large M. V. Chizhov}

\vspace{.5cm}

{\em
CERN, Geneva, Switzerland and Sofia University, Sofia, Bulgaria}

\end{center}
\vfill

\begin{abstract}
   We show that the hadron vector resonances are described by fields
transforming according to different inequivalent representations of
the Lorentz group: $(1/2,1/2)$ and $(1,0)+(0,1)$. The vector representation
$(1/2,1/2)$ is well studied and corresponds to the gauge fields.
On the other hand, the chiral representations $(1,0)$ and $(0,1)$ are
described by the second rank antisymmetric tensor fields, for which
interaction theory has not yet been constructed.
In the framework of the phenomenological Nambu--Jona-Lasinio approach we have
introduced and used all these fields for a description of the vector 
resonances. A new mass relation between 
low-lying hadron vector and axial-vector resonances is obtained. 
This relation is in agreement with the present experimental data.
\end{abstract}

\vfill

\newpage

\section{Introduction}
The bound states of a quark and an antiquark are characterized by
the quantum numbers $J^{PC}$, where $J$ is the total angular momentum,
$P$ is the parity and $C$ is the charge conjugation. There exist three
types of different quantum numbers for the known vector mesons~\cite{PDG}.
They are $1^{--}$, $1^{++}$ and $1^{+-}$. For instance, the first
quantum numbers can be assigned to the $\rho$ and $\omega$, $\phi$
vector mesons with isospin $I=1$ and $I=0$. 
The second and the third quantum numbers are assigned to
the axial-vector mesons $a_1$, $f_1$ and $b_1$, $h_1$, correspondingly.
Note that we have two different types of axial-vector mesons
and the key point is the difference between the last two assignments for these.

Let us consider the extended
Nambu--Jona-Lasinio (NJL)~\cite{rew} models.
In such models the Lagrangian contains only the quark fields, while
the spontaneous symmetry breaking and the hadron states
are generated dynamically by the model itself. The mesonic states
arise as excitations of quark--antiquark pairs and that defines their
interactions with the quarks. 

In the relativistic theory the symmetry group is the Lorentz group $O(3,1)$, 
which is isomorphic to the direct product of the two spatial rotation groups 
$O(3)\times O(3)$. 
The lowest representations of the $O(3,1)$ group,
which can be used as building blocks for the construction of 
higher spin representations, are chiral fundamental spinors
transforming according to two inequivalent representations
$(1/2,0)$ and $(0,1/2)$. 
They correspond to quarks with different chiralities.
There exists two possibilities to construct the meson fields with the spin 1.

The vector representation $(1/2,1/2)$ arises from the product between 
the left $(1/2,0)$ and the right $(0,1/2)$ fundamental spinors.
There are the vector
$\overline{\psi}\gamma_\mu\psi$ and the axial-vector
$\overline{\psi}\gamma_\mu\gamma^5\psi$ bilinear forms
of the quark spinor fields
with quantum numbers $1^{--}$ and $1^{++}$,
which couple to the vector and axial-vector mesons
respectively. They have gauge-like interactions with the quarks      
and can be described by the gauge vector $V_\mu$ and
axial-vector $A_\mu$ fields.

To describe the mesons with the quantum numbers $1^{+-}$ we must
use other inequivalent chiral representations $(1,0)$ and $(0,1)$,
which can be constructed, if one uses only the product either  
of the left $(1/2,0)$ or of the right $(0,1/2)$ fundamental spinors.
These mesonic states correspond to the bilinear form 
$\overline{\psi}\sigma_{\mu\nu}\psi$, and are
described by the antisymmetric tensor field $T_{\mu\nu}$, 
rather than by the gauge fields.
The antisymmetric tensor field $T_{\mu\nu}$ have six independent
components: three-vector and three-axial-vector, and can be decomposed 
in the axial-vector  
$B_\mu=\frac{1}{2}\epsilon_{\nu\mu\al\bt}\hat{\partial}_\nu T_{\al\bt}$
and the vector $R_\mu=\hat{\partial}_\nu T_{\nu\mu}$~\cite{TNJL},
where $i\hat{\partial}_\mu=i\partial_\mu/\sqrt{-\partial^2}$ is a
dimensionless unit vector\footnote{We assume that repeated indices are summed
in all cases.}.
Each of the vectors $B_\mu$ and $R_\mu$ has three independent components
due to the antisymmetric property of $T_{\mu\nu}$.
Besides the axial-vector mesons $B_\mu$
with quantum numbers $1^{+-}$, there are additional vector mesons $R_\mu$
with quantum numbers $1^{--}$, like those of the
gauge mesons $V_\mu$, but having different coupling to the quarks.

These excitations {\it were missed} in the NJL model and they are not 
considered as real particles at the present time. 
In this paper we shall show how
we can use the newly introduced quasi-particles in the NJL framework.
To demonstrate this idea we deal with the one-flavour NJL model only.

\section{The effective Lagrangian.}
One of the most important symmetries of the real world and QCD,
which holds in the NJL model, is chiral symmetry.
Following the classical paper
\cite{njl1} we require that the primary fermion interaction must be
invariant under $\gamma^5$- and ordinary phase transformations:
\be
\psi \to \exp[i \al \gamma^5]~\psi,~~~~~~~~~~
\bar{\psi} \to \bar{\psi}~\exp[i \al \gamma^5],
\label{chiral}
\ee
\be
\psi \to \exp[i \al]~\psi,~~~~~~~~~~~~~
\bar{\psi} \to \bar{\psi}~\exp[- i \al],
\label{usual}
\ee
where $\al$ is a constant and $\psi$ is the Dirac spinor corresponding to 
a quark field. We restrict ourselves to the consideration of quark-antiquark
bound state formations as real particles. These states are explicitly
invariant under transformations (\ref{usual}). 

Whilst the Dirac spinor has four components one can construct 16
independent bilinear forms in quark-antiquark channel:
$\bar{\psi}\psi$, $\bar{\psi}\gamma^5\psi$, $\bar{\psi}\gamma_\mu\psi$, 
$\bar{\psi}\gamma_\mu\gamma^5\psi$ and $\bar{\psi}\sg_{\mu\nu}\psi$.
Under the Lorentz group they transform as scalar, pseudoscalar, vector, 
axial-vector and antisymmetric tensor, correspondingly. To deal
with the chiral properties of these bilinear forms it is useful to
define chiral currents:
${\cal V}_\mu=\bar{\psi}\gamma_\mu\psi$, 
${\cal A}_\mu=\bar{\psi}\gamma_\mu\gamma^5\psi$, 
${\cal S}^\pm=\bar{\psi}(1\pm\gamma^5)\psi$ and
${\cal T}^\pm_{\mu\nu}=\bar{\psi}\sg_{\mu\nu} (1\pm\gamma^5)\psi$.
The vector ${\cal V}_\mu$ and axial-vector ${\cal A}_\mu$ currents 
obviously satisfy the chiral invariance.
The last two terms transform under (\ref{chiral}) as follows: 
\be
{\cal S}^\pm \to \exp[\pm 2 i \al]~{\cal S}^\pm,~~~~~~~~~~
{\cal T}^\pm_{\mu\nu} \to \exp[\pm 2 i \al]~{\cal T}^\pm_{\mu\nu}.
\ee

Now we can construct the chiral invariant Lagrangian choosing
scalar ${\cal S}^\pm$ and tensor ${\cal T}^\pm_{\mu\nu}$
current-current interactions with opposite chiralities and quadratic
forms of ${\cal V}_\mu$ and ${\cal A}_\mu$ interactions. The former is
the primary interaction in the original work \cite{njl1} of Nambu and
Jona-Lasinio.  The latter interaction
is used in the extensions of the NJL model to
achieve a sufficient attractive force in the axial-vector channel
\cite{eguchi}.  What about the tensor interactions? 
It is easy to check that its Lorentz invariant form
\be
{\cal T}^+_{\mu\nu}~{\cal T}^-_{\mu\nu} =
(\bar{\psi}\sg_{\mu\nu}\psi)^2-
(\bar{\psi}\sg_{\mu\nu}\gamma^5\psi)^2
\equiv 0
\ee
exactly equals zero, because ${\cal T}^+_{\mu\nu}$ and
${\cal T}^-_{\mu\nu}$ belong to different irreducible representations
of the Lorentz group, namely $(1,0)$ and $(0,1)$. 
It is therefore impossible to include tensor excitations in a local
one-flavour NJL model. 
Possible minimal extension has been proposed in~\cite{TNJL}.
The effective four-fermion Lagrangian has the form
\begin{eqnarray}
{\cal L}_{int} = &+&G_S~[(\bar{\psi}\psi)^2-(\bar{\psi}\gamma^5\psi)^2]
~-~G_V~(\bar{\psi}\gamma_\mu\psi)^2
-~G_A~(\bar{\psi}\gamma_\mu\gamma^5\psi)^2
\nonumber \\
&-&G_T~[\hat{\partial}_\mu(\bar{\psi}\sg_{\mu\ld}\psi)\cdot 
\hat{\partial}_\nu(\bar{\psi}\sg_{\nu\ld}\psi)-
\hat{\partial}_\mu(\bar{\psi}\sg_{\mu\ld}\gamma^5\psi)\cdot 
\hat{\partial}_\nu(\bar{\psi}\sg_{\nu\ld}\gamma^5\psi)].
\label{eff}
\end{eqnarray}

\section{The collective states.}

The nonlinear current-current interactions (\ref{eff}) can be linearized 
by means of the formula
\be
\exp[-\frac{i}{2} J{\cal K}^{-1}J]=
C\int [{\rm d}\varphi] \exp[iJ\varphi +\frac{i}{2}\varphi{\cal K}\varphi],
\label{path}
\ee
where auxiliary fields $\varphi$ will play the role of collective meson 
states. Then the initial Lagrangian takes the form
\begin{eqnarray}
{\cal L}_{init} &=&
i~\bar{\psi}\partial\bk\psi + g_S~\bar{\psi}\psi~S - \frac{g_S^2}{4G_S}~S^2 
+i~g_P~\bar{\psi}\gamma^5\psi~P  - \frac{g_P^2}{4G_P}~P^2
\nonumber\\
&+& g_V~\bar{\psi}\gamma_\mu\psi~V_\mu + \frac{g_V^2}{4G_V}~V_\mu^2
+g_A~\bar{\psi}\gamma_\mu\gamma^5\psi~A_\mu + \frac{g_A^2}{4G_A}~A_\mu^2
\nonumber\\
&+& g_R~\hat{\partial}_\nu(\bar{\psi}\sg_{\mu\nu}\psi)~R_\mu
+ \frac{g_R^2}{4G_R}R_\mu^2
+ i~g_B~\hat{\partial}_\nu(\bar{\psi}\sg_{\mu\nu}\gamma^5\psi)~B_\mu 
+ \frac{g_B^2}{4G_B}B_\mu^2.
\label{int}
\end{eqnarray}
Here we have introduced all possible low-lying collective states.
They are scalar $S$, pseudoscalar $P$, two vectors $V_\mu$, $R_\mu$
and two axial-vectors $A_\mu$, $B_\mu$ with the following quantum numbers: 
\begin{center}
\begin{tabular}{l|c|c|c|c|c|c}
meson fields:    & $S$ & $P$ & $V_\mu$ & $A_\mu$ & $R_\mu$ & $B_\mu$\\
\hline
quantum numbers: & $0^{++}$ & $0^{-+}$ & $1^{--}$ & $1^{++}$ &
$1^{--}$ & $1^{+-}$
\end{tabular}
\end{center}
Let us note that there are two different vector mesons $V_\mu$ and 
$R_\mu$ with the same quantum numbers. Therefore, the physical states 
of such mesons can be linear combinations of them\footnote{It means, 
in particular, that the $\rho$ meson can have both gauge and anomalous 
tensor couplings with the quarks~\cite{rho},
while the axial-vector mesons with quantum numbers $1^{++}$
have only gauge interactions and
the axial-vector mesons with quantum numbers $1^{+-}$
have only tensor interactions.}.

Integration over the quark field $\psi$ leads to an effective Lagrangian
for the meson fields with proper induced kinetic and mass terms.
Various interactions among the meson fields also arise~\cite{TNJL}.
Here we are only interested in a mass matrix.

\section{The mass matrix.}
It is well known that after spontaneous symmetry breaking, when the quark 
field acquires mass $m$, we get the scalar meson with the mass $2m$ and
the massless pseudoscalar meson. Due to interactions with the scalar field
$S$ the mass terms for the vector and axial-vector mesons are also modified.
The final Lagrangian for the mass terms reads
\begin{eqnarray}
{\cal L}_M =
 {M^2_V \over 2} V^2_\mu 
+ \sqrt{\frac{3}{2}} m V_{\mu\nu} \hat{R}_{\mu\nu}
+ {M^2_T - 6m^2 \over 2} R^2_\mu
+{M^2_A + 6m^2 \over 2} A^2_\mu + {M^2_T + 6m^2 \over 2} B^2_\mu,
\label{mass}
\end{eqnarray}
where $V_{\mu\nu}=\partial_\mu V_\nu - \partial_\nu V_\mu$ and
$\hat{R}_{\mu\nu}=\hat{\partial}_\mu R_\nu -\hat{\partial}_\nu R_\mu$.
Here $M_V$, $M_A$, $M_T$ and $m$ masses can be independent. But if we
believe that the effective four-fermion interactions of the quarks
could originate in QCD by gluon exchange in the $1/N_c$ limit, one obtains 
$M_V=M_A$ \cite{rew}. This reduces the number of unknown parameters to 
three. Therefore, we can derive one relation 
for four physical masses.

As it is expected, the vector mesons with the same quantum numbers are mixed
\be
{\cal M}^2(q^2) = \left(
\begin{array}{cc}
M^2_V & \sqrt{6 m^2 q^2} \\
\sqrt{6 m^2 q^2} & M^2_T - 6m^2
\end{array} \right).
\label{mix}
\ee
Let us note that it is dynamical mixing, because the mass matrix 
contains explicitly momentum $q^2$. Therefore, the mixing angle
depends on the momenta.

As long as the isospin triplets consist of $up$ and $down$
quarks with approximately the same constituent
masses we can apply this one-flavour
model to the charged mesons $\rho$, $a_1$, $b_1$ and $\rho'$
in order to avoid mixing with $s$ quark for $I=0$ multiplets.
In this case $m^2_{a_1}\equiv M^2_A + 6m^2$, $m^2_{b_1}\equiv M^2_T + 6m^2$,
and $m^2_\rho=\lambda_1(m^2_\rho)$,
$m^2_{\rho'}=\lambda_2(m^2_{\rho'})$ 
are solutions of the quadratic equation:
\be
\lambda^2-(m^2_{a_1}+m^2_{b_1}-12m^2)~\lambda+
(m^2_{a_1}-6m^2)(m^2_{b_1}-12m^2) = 0.
\label{roots}
\ee
Using the Vi\`{e}te theorem we get immediately from (\ref{roots}) 
the following useful relationships:
\be
\left\{
\begin{array}{ccl}
m^2_{\rho'}+m^2_\rho& =& m^2_{a_1}+m^2_{b_1}-12m^2,\\
m^2_{\rho'}~m^2_\rho & =& (m^2_{a_1}-6m^2)(m^2_{b_1}-12m^2),
\end{array}
\right.
\ee
or
\be
\newlength{\lll}
\setlength{\lll}{8cm}
\addtolength{\lll}{-2\fboxsep}
\addtolength{\lll}{-2\fboxrule}
\noindent
\fbox{%
\begin{minipage}{\lll}
\[
m^2_{b_1}=\frac{m^4_{\rho'}+m^4_\rho-m^4_{a_1}} 
{m^2_{\rho'}+m^2_\rho-m^2_{a_1}}.
\]
\label{result}
\end{minipage}}
\ee
The last equation is the main result of our work, which can be compared
with experiment.

As long as the masses of $\rho$ and $b_1$ mesons are known with better
precision than the $\rho'$ or $a_1$ masses, we take them as input
parameters~\cite{PDG}: $m^{exp}_\rho=769.9\pm 0.8$ MeV and 
$m^{exp}_{b_1}=1229.5\pm 3.2$ MeV.
Then we can compare the predicted $m_{\rho'}$~--~$m_{a_1}$ relation
($3\sigma$ allowed region between curves in Fig.~1) with 
the experimental data. For this purpose we have shown in Fig.~1 
the combined result for the charged $\rho' \equiv \rho(1450)$ mass 
measurements~\cite{rho_mass} and the PDG result~\cite{PDG} 
for $a_1$ mass with $3\sigma$ allowed intervals.
The last high-precision measurement of $a_1$ mass~\cite{Asner} is also
shown.

\begin{figure}[h]
\begin{center}
\epsfig{file=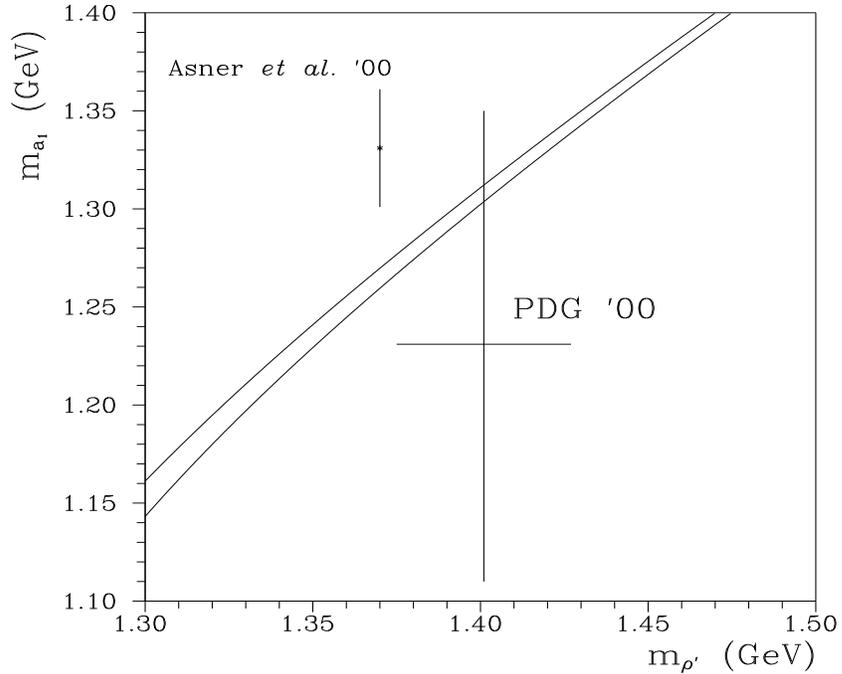,height=8.9cm,width=11cm}
\caption{The $3\sigma$ allowed region for $\rho'$ and $a_1$ masses 
versus the experimental data.}
\end{center} 
\end{figure}

We can conclude that there is reasonable agreement between the prediction
(\ref{result}) and the experimental data. 
Quantitatively, the higher value of $a_1$ mass $m_{a_1} = 1307\pm 40$ MeV 
is favoured (in comparison with PDG), 
if the following $\rho'$ mass value 
$m_{\rho'} = 1401\pm 26$ MeV~\cite{rho_mass,rhoprime_last} 
is accepted. The latter is confirmed by the latest experimental data. The
corresponding quark mass is $m=235.5\pm 3.4$ MeV. It leads to the prediction
for the mass of the sigma meson $m_\sigma=2m=471\pm 7$ MeV that is also
in excellent concordance with the recent experiment~\cite{sigma} 
$m^{exp}_\sigma= 478\pm 29$ MeV.

\pagebreak[2]

\end{document}